# Revealing the intrinsic superconducting gap anisotropy in surface-neutralized BaFe$_2$(As$_{0.7}$P$_{0.3}$)$_2$

Ziming Xin[1], Yudi Wang[1], Cong Cai[1], Zhengguo Wang[1], Lei Chen[1], Tingting Han[1], and Yan Zhang[1, 2, *]

[1]International Centre for Quantum Materials, School of Physics, Peking University, Beijing 100871, China

[2]Collaborative Innovation Centre of Quantum Matter, Beijing 100871, China

**Alkaline-earth iron arsenide (122) is one of the most studied families of iron-based superconductors, especially for angle-resolved photoemission spectroscopy. While extensive photoemission results have been obtained, the surface complexity of 122 caused by its charge-non-neutral surface is rarely considered. Here, we show that the surface of 122 can be neutralized by potassium deposition. In potassium-coated BaFe$_2$(As$_{0.7}$P$_{0.3}$)$_2$, the surface-induced spectral broadening is strongly suppressed, and hence the coherent spectra that reflect the intrinsic bulk electronic state recover. This enables the measuring of superconducting gap with unpreceded precision. The result shows the existence of two pairing channels. While the gap anisotropy on the outer hole/electron pockets can be well fitted using an s$_\pm$ gap function, the gap anisotropy on the inner hole/electron shows a clear deviation. Our results provide quantitative constraints for refining theoretical models and also demonstrate an experimental method for revealing the intrinsic electronic properties of 122 in future studies.**

## Introduction

After the discovery of superconductivity in $LaO_{1-x}F_xFeAs$ (1111), many families of iron-based superconductor have been found, including iron-selenide (11), alkaline iron arsenide (111), alkaline-earth iron arsenide (122), etc[1, 2]. Among them, the 122 family is the most studied family of iron-based superconductors, due to its high sample quality, high superconducting transition temperature ($T_c$), tunable carrier density, and diversity of compounds with different chemical substitutions. However, aside from these advantages, the lattice structure of 122 contains a single alkaline-earth-metal plane and hence has no charge-neutral cleavage surface (Fig. 1a,b). In $BaFe_2(As_{1-x}P_x)_2$, for example, half Ba ions are removed at the cleaved surface, and the residual Ba ions distribute inhomogeneously, forming various surface terminations[3-7]. While the alkaline-earth-metal-terminated and arsenic-terminated surfaces have been observed by scanning tunneling microscopy (STM)[4-7], the alkaline-earth-metal-deficient surface also exists with a reconstruction of alkaline-earth-metal atoms with 1×2 and √2×√2 periods[6, 7]. In bulk materials, one alkaline-earth-metal plane donates two electrons to the two nearest Fe-As/P planes. However, at the surface, the number of charge that transfers to the topmost Fe-As/P plane varies strongly depending on the concentration of alkaline-earth-metal atoms, leading to a charge-inhomogeneous surface for the 122 iron-based superconductors.

Angle-resolved photoemission spectroscopy (ARPES) is a powerful technique that measures the electronic structure of material in momentum space. In the studies of iron-based superconductors, ARPES plays an important role in determining band structure, Fermi surface topology, superconducting gap anisotropy, etc[8, 9]. However, as a surface-sensitive technique, ARPES is very sensitive to the cleavage surface of material. For example, in 1111, the separation of bulk and surface states has been observed, which was attributed to the charge-non-neutral cleavage surface of 1111[10-12]. However, for the most studied family of iron-based superconductors, the 122 family, the complexity caused by its charge-non-neutral surface is rarely considered. While few theories and experiments show the possible existence of surface state in 122[3,13], the photoemission spectra taken in 122 are generally much broader[13-20] than that taken in other families of iron-based superconductors[21-25]. Such broadening behavior and its origin have not been well understood so far.

Here, we report the measurement of gap anisotropy in an optimal-doped 122 compound BaFe$_2$(As$_{0.7}$P$_{0.3}$)$_2$ utilizing ARPES and in-situ potassium deposition. We find that by depositing a small amount of potassium on the sample surface, the spectra become sharp and coherent, which allows us to measure the superconducting gap anisotropy on all Fermi surface sheets with unpreceded precision. We show that the obtained gap anisotropy cannot be fitted using a single |cos$k_x$cos$k_y$| gap function, but could be explained by the nesting- and orbital-selectivity of the superconducting pairing. Our detailed and precise gap measurement provide crucial clues for uncovering the paring mechanism of iron-based superconductors. It also implies that the surface complexity of 122 need to be seriously considered. The potassium deposition can be used as a practical experiment method for revealing the intrinsic electron structure and gap anisotropy of 122 iron-based superconductors in the future studies.

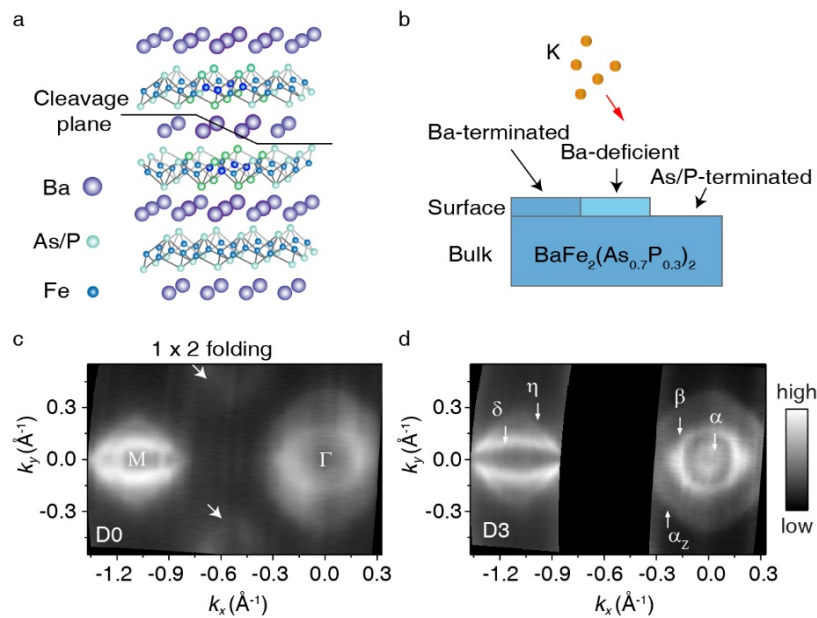

**Fig. 1 Surface complexity of BaFe$_2$(As$_{0.7}$P$_{0.3}$)$_2$** **a** Schematic drawing of the lattice structure of BaFe$_2$(As$_{0.7}$P$_{0.3}$)$_2$. The black solid line illustrates the position of a possible cleavage plane. **b** Schematic drawing of the cleavage surface and potassium deposition. **c** Fermi surface mapping taken in the pristine sample (D0). $k_x$ and $k_y$ are the momenta of electrons along x and y directions. **d** is the same as **c** but taken in the doped sample (D3) after three times potassium deposition.

## Results

**Evolution of electronic structure in potassium-coated BaFe$_2$(As$_{0.7}$P$_{0.3}$)$_2$.** The Fermi surface mappings taken in the pristine and doped samples are compared in Fig. 1c,d. The photoemission spectra distribute broadly in the pristine sample (D0) and shadow Fermi pockets are observed at the Brillouin zone boundary (X), indicating a 1×2 reconstruction of the cleavage surface. With potassium deposition, the sharpening of photoemission spectra is obvious. Two hole pockets ($\alpha$ and $\beta$) and two electron pockets ($\delta$ and $\eta$) are now clearly resolved at the Brillouin zone center ($\Gamma$) and zone corner (M). The sharpness of photoemission spectra taken in the doped sample transcends most previous ARPES studies on 122, especially for the Fermi surface mapping taken around the M point[8, 9, 14-17]. Note that, the large hole pocket ($\alpha_Z$) is a shadow Fermi pocket that is folded from the Z point due to the finite $k_z$ resolution of ARPES[17].

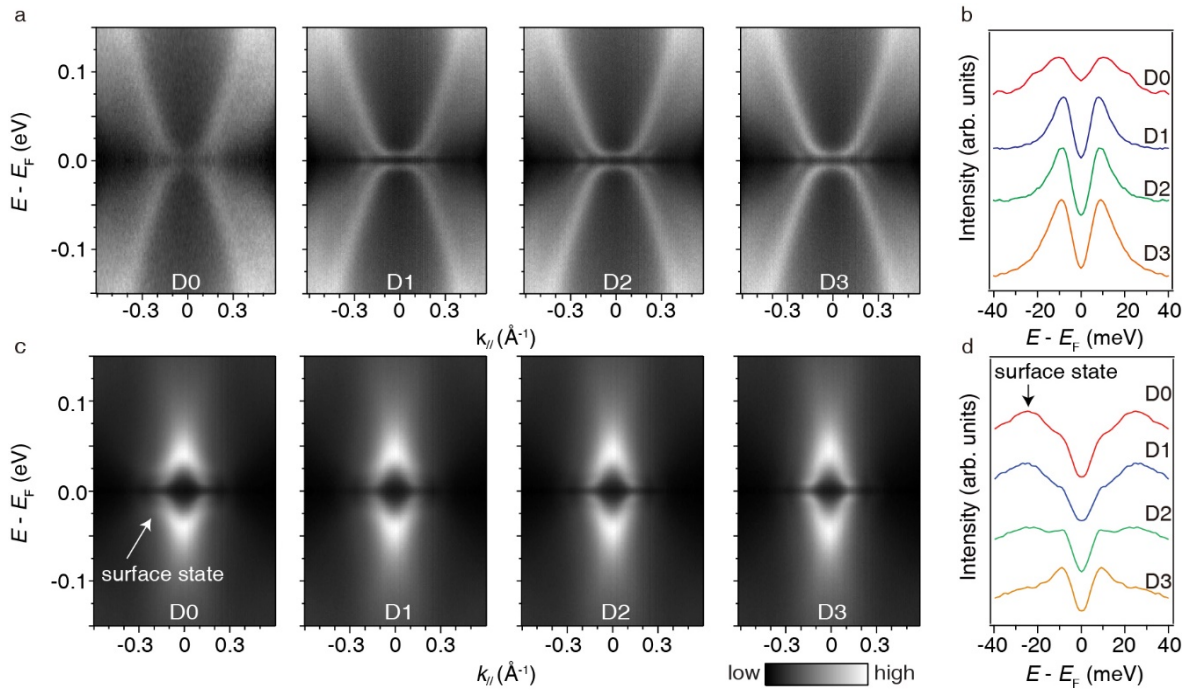

**Fig. 2 Band evolution in potassium-coated BaFe$_2$(As$_{0.7}$P$_{0.3}$)$_2$. a** Doping dependence of the symmetrized energy-momentum cuts taken around the $\Gamma$ point. $E - E_F$ represents the binding energy of electrons. $E$ is the kinetic energy of electrons and $E_F$ is the Fermi energy. $k_{//}$ is the in-plane momentum of electrons. **b** Doping dependence of the symmetrized energy distribution curves (EDCs) taken at the Fermi crossing ($k_F$) of the $\alpha$ hole band. **c** and **d** are the same as **a** and **b** but taken around the M point. The EDCs are taken at the $k_F$ of the $\delta$ electron band. The deposition sequence is denoted using Dn (n is the doping times).

To further show the effect of potassium deposition, we take the energy-momentum cuts across the hole/electron pockets and plot their doping dependence in Fig. 2. Around the Γ point, the spectra become sharper with potassium deposition (Fig. 2a), and meanwhile the superconducting peaks become more coherent (Fig. 2b). Around the M point, a surface band could be observed at around -20 meV as pointed out by the white arrow (Fig. 2c). With potassium deposition, this surface band weakens and eventually diminishes. As a result, the superconducting coherent peak is clearly resolved (Fig. 2d). It should be noted that, we could not resolve any change of Fermi crossings ($k_F$s) and gap magnitudes (Δ), which indicates that the total potassium coverage is very low. Based on our energy and momentum resolutions, we could set an upper bound of total potassium coverage to be ~0.015 ML. For each doping step, the increment of potassium coverage is estimated to be below 0.005 ML. Such small amount of potassium is sufficient to improve the spectral quality while having little influence on band structure and superconductivity.

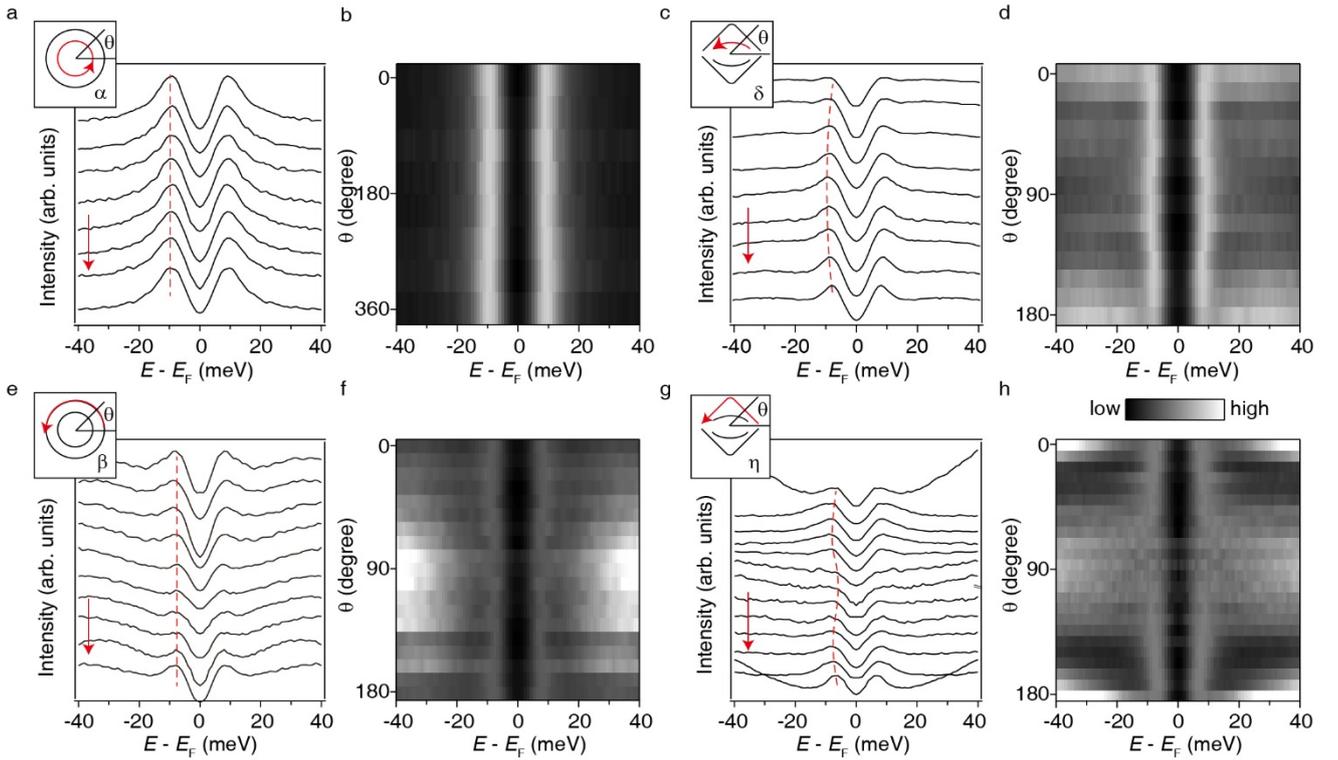

**Fig. 3 Angular distributions of superconducting gap on Fermi surface sheets**. **a** Angle dependence of the symmetrized energy distribution curves (EDCs) taken on the α hole pocket. The red dashed line guides the eyes to the gap anisotropy. $E - E_F$ represents the binding energy of electrons. $E$ is the kinetic energy of electrons and $E_F$ is the Fermi energy.

The red arrow illustrates the direction of the data presentation. **b** The merged image of the symmetrized EDCs for better visualizing the gap anisotropy. The polar angle (θ) is defined in **a**. **c**, **d**, **e**, **f**, **g**, and **h** are the same as **a** and **b**, but taken on the β, δ, and η pockets respectively. All data were taken in the D3 sample at 8 K.

**Superconducting gap anisotropy in potassium-coated BaFe$_2$(As$_{0.7}$P$_{0.3}$)$_2$.** Being able to resolve the intrinsic ARPES spectra is critical for the quantitative spectra analysis. The sharpness of spectra taken in potassium-coated BaFe$_2$(As$_{0.7}$P$_{0.3}$)$_2$ is now comparable with that taken in 11 and 111 iron-based superconductors[21-25] (Supplementary Fig. 1). On one hand, the effective mass and lifetime of quasiparticles could be determined more accurately, which is important for studying the quantum critical phenomena of BaFe$_2$(As$_{1-x}$P$_x$)$_2$ [26, 27]. On the other hand, the superconducting coherent peaks are well defined on all Fermi surface sheets, which allows us to measure the superconducting gap anisotropy with unpreceded precision. The results are shown in Fig. 3. The superconducting gap is nearly isotropic on the α and β hole pockets ($\Delta_\alpha$ and $\Delta_\beta$) (Fig. 3a-d), but shows moderate anisotropy on the δ and η electron pockets ($\Delta_\delta$ and $\Delta_\eta$) (Fig. 3e-h). For $\Delta_\delta$, the gap reaches maximum at 90°, while for $\Delta_\eta$, the gap maxima locate at 45° and 135°.

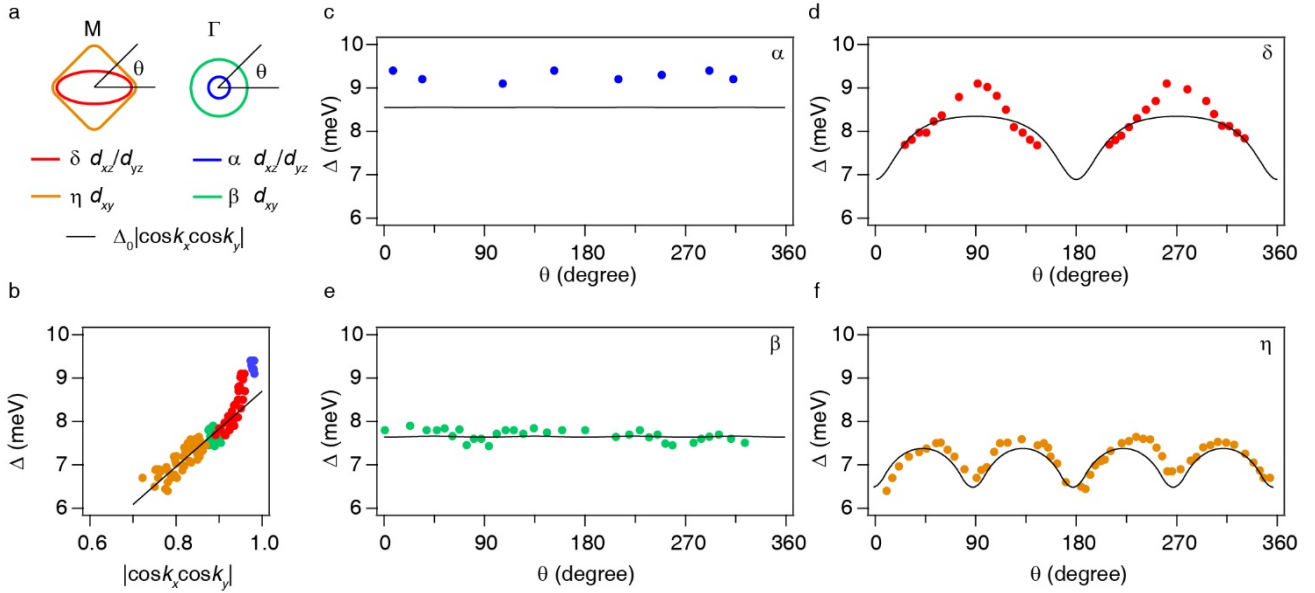

**Fig.4 Superconducting gap fitting using an s$_\pm$ gap function. a** Schematic drawing of the Fermi surface of BaFe$_2$(As$_{0.7}$P$_{0.3}$)$_2$. The orbital characters of Fermi surface are illustrated using different colors[14, 17, 18, 28, 29]. **b** Plot of the superconducting gap (Δ) as a function of |cos$k_x$cos$k_y$|. **c** Angular distribution of the superconducting gap on the α pocket. The polar

angle ($\theta$) is defined in **a**. **d**, **e**, and **f** are the same as **c** but taken on the $\delta$, $\beta$, and $\eta$ pockets respectively. The black solid lines in all panels are the best fit using the gap function $\Delta_0|\cos k_x \cos k_y|$ with $\Delta_0$ = 8.7 meV.

Figure 4 summarizes the angular distributions of superconducting gap on all Fermi surface sheets. It is commonly believed that the pairing symmetry of 122 is an $s_\pm$-wave[30-33] and the superconducting gap anisotropy can be fitted using a $\Delta_0|\cos k_x \cos k_y|$ gap function[16, 18-20]. Here, our detailed and precise gap measurements allow us to test the validity of $s_\pm$ gap function quantitatively. We fitted the data using the experimental determined Fermi surface (Fig. 4a) and summarized the fitting results in Fig. 4b-f. For the superconducting gap anisotropies on the $\beta$ and $\eta$ pockets, the four-fold symmetry of $\Delta_\eta$ and the relative gap magnitudes of $\Delta_\beta$ and $\Delta_\eta$ can be well described by the $s_\pm$ gap function with $\Delta_0$ = 8.7 meV. However, for the $\alpha$ and $\delta$ pockets, the superconducting gap clearly deviates from the $s_\pm$ gap function with a larger gap magnitude, which indicates a stronger superconducting pairing on the $\alpha$ and $\delta$ pockets.

## Discussion

We first discuss how the potassium atoms play roles on the surface of $BaFe_2(As_{1-x}P_x)_2$. There are several possible scenarios. First, electrons transfer from the potassium atoms to the sample surface, which neutralizes the charge-non-neutral surface, leading to a suppression of the surface broadening effect. Second, the potassium atoms could act as a catalyzer which causes the redistribution of alkaline-earth metal atoms on the sample surface in a more homogeneous way. Third, the potassium atoms scatter electrons at the sample surface. The surface electronic states then turn into an incoherent and continuous background that is inconspicuous in photoemission spectra. To understand the mechanism of potassium deposition, further experimental and theoretical studies are required. Nevertheless, our results highlight the surface complexity of 122 iron-based superconductors. This implies that previous photoemission results taken on 122 should be carefully revisited.

For example, ARPES observed double peak features in $Ba_{1-x}K_xFe_2As_2$, whose origin remains contraversal[34-37]. Some attribute it to the band degeneracy[34, 35], while others believe it is originated from the electron-bosonic coupling[36] or in-gap impurity state[37]. Moreover, ARPES data taken on $BaFe_2As_2$, $SrFe_2As_2$, $CaFe_2As_2$, etc[38-40] show complex band structures. The number of bands observed by ARPES is inconsistent with the band calculations. Here, we show that the surface complexity might explain these controversial behaviors of 122 iron-based superconductors. The possible existence of surface-related features could be easily verified using potassium deposition. For $BaFe_2(As_{1-x}P_x)_2$, its nodal location is still under debates[18, 28, 29, 41]. We measured the $k_z$ dependence of superconducting gap in the D3 sample (Supplementary Fig. 2). While our high-quality data confirm the existence of gap nodes on the hole pockets around the Z point[18], no gap node is resolved on the electron pockets. Although we cannot fully exclude the possible existence of nodal loop on electron pockets due to our finite $k_z$ resolution, we show that the data quality can be significantly improved by potassium deposition especially for the gap measurement on the electron pockets. With a more bulk-sensitive and comprehensive photon energy dependent experiment, it can be determined whether the nodal loop exists or not.

We now turn to the in-plane superconducting gap anisotropy, which has been studied by ARPES for many iron-based superconductors. For 111 and 11 compounds, such as LiFeAs, $NaFe_{1-x}Co_xAs$ and FeSe, the Fermi pockets are small and the $T_c$s are relatively low, making the gap function fitting unreliable and unable to provide effective information[8, 23-25]. For 122 compounds, due to the surface broadening effect, the superconducting gap anisotropy, especially the gap anisotropy on the electron pockets, has never been resolved clearly. Here, we achieve an accurate and reliable gap measurement of 122 using potassium-coated $BaFe_2(As_{0.7}P_{0.3})_2$. Taking all advantages of 122, such as large Fermi pocket size and high $T_c$, our data provide an ideal touchstone for testing the theoretical models of iron-based superconductors.

First, we show that the superconducting pairing cannot be described by a single $s_\pm$ pairing channel, but instead consists of at least two pairing channels. One pairing channel follows the $s_\pm$ pairing symmetry and dominates the superconducting pairing on the $\beta$ and $\eta$ pockets. The other pairing channel could be described as a second $s_\pm$ pairing channel with a larger gap pre-factor. It dominates the superconducting pairing on the α and δ pockets. Our results undoubtedly show that the single gap function that normally describes a single band superconductor fails to explain the gap anisotropy of iron-based superconductors. The existence of multiple pairing channels and the interplays among them should be considered in the description of pairing interactions in iron-based superconductors.

Second, according to the theories of iron-based superconductors[30-33], the pair scattering of electrons is mediated by a large **Q** scattering that connects the Fermi pockets at the Γ and M points. Here, our gap fitting shows that two pair scattering channels connect the outer electron and hole pockets (β and η) and the inner electron and hole pockets (α and δ), respectively. On one hand, the outer hole and electron pockets are similar in size and are well nested. This is also the case for the inner hole and electron pockets. The nesting of Fermi surface naturally separates out two pair scattering channels which locate respectively on the outer and inner Fermi pockets. On the other hand, while the β and η pockets are constructed by the $d_{xy}$ orbital, the α and δ pockets are constructed by the $d_{xz}/d_{yz}$ orbitals. The intra-orbital pair scattering could also lead to a separation of two pairing channels respectively from the $d_{xz}/d_{yz}$ and $d_{xy}$ orbitals. Based on above discussions, our results indicate that the intra-orbital pair scattering between two nested Fermi pockets play a dominating role in the superconducting pairing of iron-based superconductors. As a result, the superconducting pairing becomes nesting- and orbital-selective[42, 43]. By fitting the gap anisotropy using a more complex and parameterized gap function, the detailed momentum and orbital dependence of pairing interactions could be obtained, which would help to construct an accurate and realistic superconducting pairing model for iron-based superconductors.

In summary, we report the superconducting gap measurement in potassium-coated $BaFe_2(As_{0.7}P_{0.3})_2$. We found that a small amount of potassium deposition could suppress the surface-broadening effect and help us to reveal the intrinsic electron structure of $BaFe_2(As_{0.7}P_{0.3})_2$. Our high quality and precise gap measurement distinguishes two pairing channels, which unveils the nesting- and orbital-selective nature of the superconducting pairing of iron-based superconductors. Our results show that the surface-neutralization enables the photoemission data taken in 122 to be analyzed with unpreceded precision.

The results would provide crucial clues for uncovering the pairing mechanism of iron-based superconductors.

## Methods

**Sample preparations.** High quality BaFe$_2$(As$_{0.7}$P$_{0.3}$)$_2$ single crystals were grown using self-flux method[44]. Ba$_2$As$_3$, Ba$_2$P$_3$, FeAs, FeP were starting materials, which were mixed at a molar ratio of 2.82 : 0.18 : 0.94 : 0.06, placed in an Al$_2$O$_3$ crucible, and sealed in an iron crucible. The crucible was heated at 1150 ℃ for 10 hours, and then the temperature cooled down to 900 ℃ at a rate of 1 ℃ per hour. Finally, 1 mm × 1 mm × 0.2 mm high-quality single crystal can be obtained. The $T_c$ is around 30 K as confirmed by magnetic susceptibility measurement.

**Angle-resolved photoemission spectroscopy.** The ARPES data were taken at Stanford synchrotron radiation lightsource (SSRL) beamline 5-4. The photon energy is 23 eV. The overall energy resolution is around 5 meV and the angular resolution is around 0.3°. The samples were cleaved *in-situ* and measured in vacuum better than 5×10$^{-10}$ mbar. All data were measured at 8 K. The potassium deposition was conducted in-situ using a potassium dispenser. We repeated the deposition several times. The deposition sequence is denoted using Dn (n is the doping times). The current of the potassium dispenser is ~5.4 A, and each deposition last for ~8 seconds.

## Data availability

Data that support the findings of this study are available upon reasonable request from the corresponding authors.

**Acknowledgements**

We gratefully acknowledge the experimental support by D. H. Lu and M. Hashimoto at SSRL. This work is supported by the National Natural Science Foundation of China (Grant No. 11888101), the National Key Research and Development Program of China (Grant No. 2018YFA0305602 and No. 2016YFA0301003), and the National Natural Science Foundation of China (No. 91421107 and No. 11574004). Stanford Synchrotron Radiation Lightsource is operated by the Office of Basic Energy Sciences, U.S. Department of Energy.

**Author contributions**

Y. Z. conceived and instructed the project. Z. M. X. synthesized the single crystals. Y. D. W. took the ARPES measurements with the contribution of T. T. H., C. C., Z. G. W., L. C., and Z. M. X.. Z. M. X. and Y. Z. analysed the data and wrote the paper with the input from all authors.

**Competing interests**

The authors declare no competing interests.

**Additional Information**

Correspondence and request for materials should be addressed to Y. Zhang (yzhang85@pku.edu.cn).

# Supplementary Information

# Revealing the intrinsic superconducting gap anisotropy in surface-neutralized BaFe$_2$(As$_{0.7}$P$_{0.3}$)$_2$

Xin *et al*.

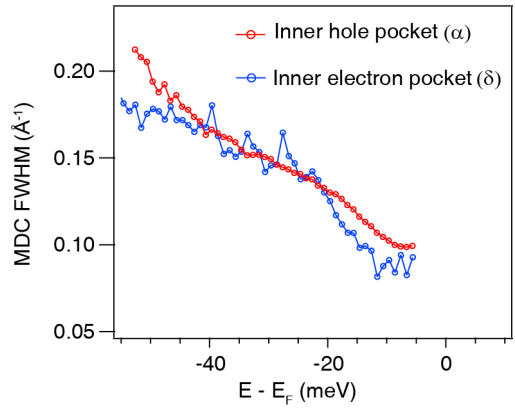

**Supplementary Figure 1 | MDC fitting on the inner hole and electron pockets.** We fit the MDCs using Lorentz peaks. The original data is shown in Fig. 2 in the main text. The data were taken on the inner hole and inner electron bands in the D3 sample. The energy dependences of MDC full-width-half-maximum (FWHM) are plotted. First, the MDC FWHM is small in comparison with most previous ARPES studies on 122 iron-based superconductors. The MDC FWHM is comparable to that measured in LiFeAs and FeSe[1,2]. Such small MDC width indicates that the spectral quality is much more improved by potassium deposition. Second, a kink-like feature could be observed at ~20meV, which is consistent with the mode coupling behaviour that has been reported in LiFeAs and $Ba_{1-x}KFe_2As_2$[1,3]. Being able to resolve the intrinsic ARPES spectra provides us a good opportunity to study the quantum critical phenomena of $BaFe_2(As_{1-x}P_x)_2$[4,5]. It is intriguing to further study how the effective mass and lifetime of quasi-particle evolves with temperature and doping in potassium-coated $BaFe_2(As_{1-x}P_x)_2$.

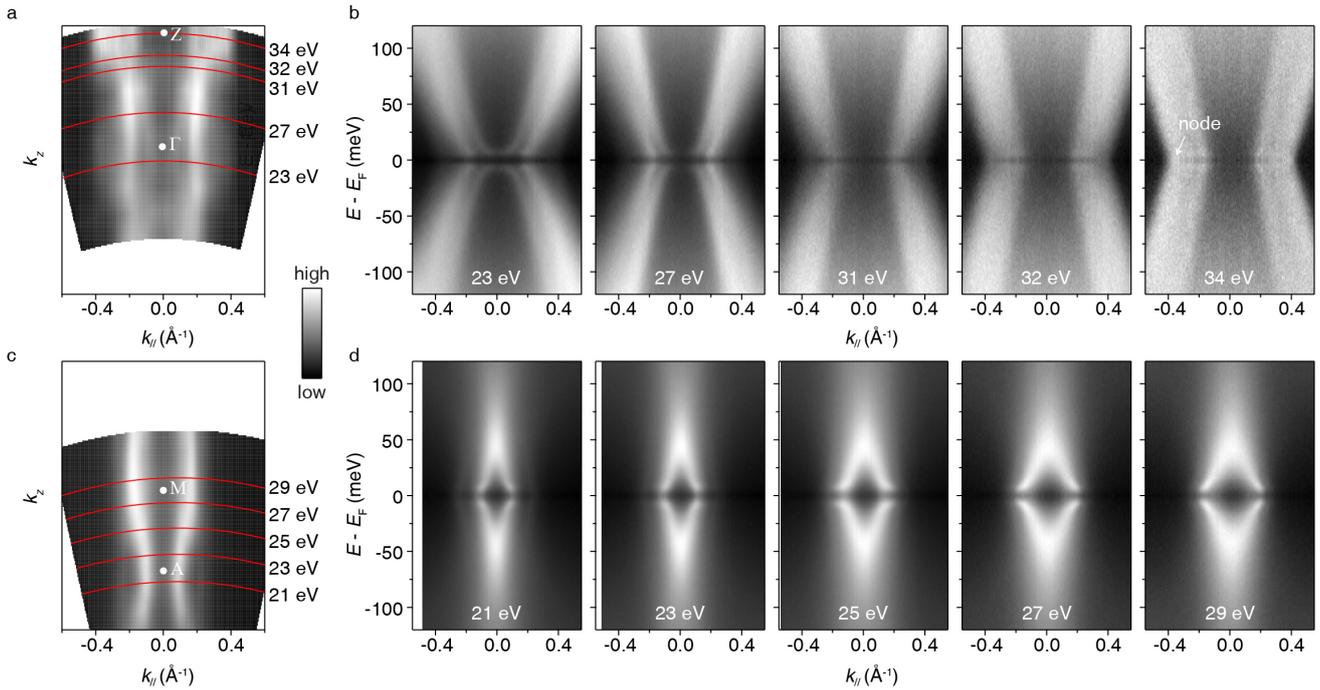

**Supplementary Figure 2 | $k_z$ dependence of superconducting gap on hole and electron pockets**. BaFe$_2$(As$_{0.7}$P$_{0.3}$)$_2$ exhibits three-dimensional electronic structure and gap nodes characteristics[6,7]. Regarding to the nodal locations, there are two possible scenarios, the gap nodes locate at the outmost hole pocket around the Z point[8] or the gap nodes locate at the tip of the outer electron pocket[9]. Here, we measured the $k_z$ dependence of superconducting gap in potassium-coated BaFe$_2$(As$_{0.7}$P$_{0.3}$)$_2$. For the hole pockets, we do observe gap nodes on the outmost hole pocket around the Z point, which is well consistent with Ref. 8. For the electron pockets, our data quality is much more improved in comparison to previous ARPES studies[7-9]. However, we did not resolve any nodes on the electron pockets. While the existence of node on the hole pocket is conclusive, for the electron pockets, we still cannot fully exclude the possible existence of nodal loop due to our finite $k_z$ resolution. More detailed and bulk-sensitive photon energy dependent data taken on potassium-coated BaFe$_2$(As$_{0.7}$P$_{0.3}$)$_2$ are necessary to determine whether nodal loop exists on electron pockets or not.

## Supplementary References